\documentclass[3p,preprint,12pt,sort&compress]{elsarticle}
\usepackage{eurosym}
\usepackage{amsmath}
\usepackage{amssymb}
\usepackage{amsfonts}
\usepackage{ulem}
\usepackage{xcolor}
\usepackage{epsfig}
\usepackage{float}

\newtheorem{definition}{Definition}

\newenvironment{proof}[1][Proof]{\noindent\textbf{#1.} }{\ \rule{0.5em}{0.5em}}

\journal{arXiv}
\begin{document}
	
\begin{frontmatter}

\title{Effect of intratumor heterogeneity in managing\\
the go-or-grow dichotomy of cancer cells:\\
a game theory modeling to understand metastasis}

\author[PUC]{Andr\'e Rocha\corref{cor1}}
\ead{andre-rocha@puc-rio.br}
\author[SGBH,UT]{Claudia Manini}
\author[BBRI]{Jos\'e I L\'opez}
\author[UPV1,IKE]{Annick Laruelle}

\address[PUC]{Department of Industrial Engineering,	Pontifical Catholic University of Rio de Janeiro,\\
	Rua Marqu\^es de S\~ao Vicente 225, G\'avea, CEP22451-900, Rio de Janeiro, RJ, Brazil.}
\address[SGBH]{Department of Pathology, San Giovanni Bosco Hospital, 10154 Turin, Italy.}
\address[UT]{Department of Sciences of Public Health and Pediatrics, University of Turin, 10124 Turin, Italy.}
\address[BBRI]{Biomarkers in Cancer Group, Biobizkaia Health Research Institute, 48903 Barakaldo, Spain.}
\address[UPV1]{Department of Economic Analysis (ANEKO), University of the Basque Country (UPV/EHU);\\ Avenida Lehendakari Aguirre, 83, E-48015 Bilbao, Spain.}
\address[IKE]{IKERBASQUE, Basque Foundation of Science, 48011, Bilbao, Spain.}

\cortext[cor1]{Corresponding author}

\begin{abstract}
	We study the effect of intratumor heterogeneity in the likelihood of 
	cancer cells moving from a primary tumor to other sites in the human body, 
	generating a metastatic process. We	model different scenarios of competition 
	between tumor cells using a static evolutionary game in which cells 
	compete for nutrients and oxygen and might choose to stay and proliferate in the 
	primary tumor or opt to a motility strategy in order to find resources in a metastatic site.
	The theoretical results found in the evolutionarily equilibrium in the mathematical model
	are in line with the empirical results observed in oncology, namely, the coexistence
	of both primary and metastatic tumors and the conditions that favor a
	metastatic process. Particularly, the model finds mathematical support for what is empirically 
	observed in punctuated and branching cancers for the specific case of clear cell renal 
	cell carcinomas: motility of cells is larger in punctuated cancers if the proportion of 
	\textit{BAP1} mutations remain below a given cell proportion threshold. 
\end{abstract}

\begin{keyword}
	Cancer\sep Intratumor heterogeneity\sep Cell motility\sep Clear cell renal cell carcinoma
\end{keyword}

\end{frontmatter}

\section{Motility of cancer cells as a go-or-grow game}

Clear cell renal cell carcinoma (CCRCC) is the most common subtype of renal
cancer and a significant subset of them pursue an aggressive clinical
course, with high mortality rates usually linked to metastatic spread. CCRCC
is a neoplasm associated to \textit{VHL} gene driver mutations and is also a
paradigm of intratumor heterogeneity where different clones develop along
the natural history of the disease. Among them, \textit{BAP1} and \textit{%
SETD2} genes mutations are associated with different clinical
aggressiveness. \textit{BAP1} gene mutations are linked to very aggressive
behavior, with early and multiple metastases and a punctuated-type tumor
evolution, whereas \textit{SETD2} gene mutations associate to a less
aggressive clinical course, with late and solitary metastases and a
branching-type evolutionary model \cite{Turajlic18}. Our key objective is to
study the relation between heterogeneity and cellular motility in CCRCC.

In a similar vein as in \cite{Basanta08} and \cite{Dwivedi23} we study the
hypothetical reason that explains why cancer cells move from a primary tumor
to other sites in the human body, generating a metastatic process. In
particular we are interested in the impact of intratumor heterogeneity. We
model different scenarios of competition between tumor cells using a static
evolutionary game in which two cells compete for nutrients and oxygen
choosing one of two possible strategies, i.e., proliferative or motile.

If the cell adopts the proliferative strategy it remains in the primary
tumor, shares the nutrients and oxygen available there with other cells, and
replicates into new tumor cells (tumor growth). Instead, the motile strategy
means to abandon the primary tumor by spreading through the bloodstream into
distant sites (metastases). This second strategy enables cells to
successfully escape the nutrient deficiency present in a given time in the
primary tumor at the expense of the energy cost of moving. In the following
game \textquotedblleft $a$\textquotedblright\ represents the amount of
nutrients and oxygen available in the metastatic site, \textquotedblleft $b$%
\textquotedblright\ indicates the same parameters available in the primary
tumor, and \textquotedblleft $c$\textquotedblright\ quantifies the energy
cost of taking the motile strategy to leave the primary tumor. The level of
nutrients and energy in a cell are quantified in terms of the amount of
adenosin-triphosphate (ATP) units.

In this context, the energy trade-off between the amount of nutrients
obtained and the cost of obtaining them determines the fitness of cells. The
fitness depends on the strategies adopted by cells. In a bilateral
encounter, if one cell adopts the proliferative strategy and the other cell
chooses the motile one, the cell adopting the proliferative strategy gets
all the nutrients in the primary tumor, \textquotedblleft $b$%
\textquotedblright , while the cell adopting the motile strategy obtains all
the nutrients in the metastatic site minus the cost of moving,
\textquotedblleft $a-c$\textquotedblright . If both cells adopt the same
strategy, nutrients and energy costs are shared. In particular, if both
cells adopt the proliferative strategy, each one gets half the nutrients in
the primary tumor, \textquotedblleft $b/2$\textquotedblright . If both cells
adopt the motile strategy each cell gets half the nutrients in the
metastatic site, \textquotedblleft $a/2$\textquotedblright , paying half of
the cost of moving, \textquotedblleft $c/2$\textquotedblright , leading to
an overall payoff of \textquotedblleft $a/2-c/2$\textquotedblright . These
contingencies are summarized in the following payoff matrix: 
\begin{equation}
\begin{tabular}{c|cc}
& Motile & Proliferative \\ \hline
Motile & $a/2-c/2$ & $a-c$ \\ 
Proliferative & $b$ & $b/2$%
\end{tabular}
\label{Matrix}
\end{equation}

We assume the following fitness ordering: $a-c>b>a/2-c/2>b/2$. That is, the
largest fitness is obtained by a metastasizing cell (a cell adopting the
motile strategy) if the opponent cell decides to remain in the primary tumor
(the opponent adopts the proliferative strategy). In this case, the cost of
moving is largely compensated by the amount of nutrients present in the
metastatic site, and we have $a-c>b$. The second best option for a cell is
to adopt the proliferative strategy if the other one takes the motile
strategy and moves away. In this case the cell obtains the whole amount of
nutrients present in the primary tumor because the intercellular competition
no longer exists. The amount of nutrients in the primary tumor is larger
than the amount obtained if both cells decide to move, $b>a/2-c/2$. Finally,
the worst alternative is when both cells choose the proliferative strategy
remaining at the primary tumor and sharing the nutrient scarcity, as we have 
$a/2-c/2>b/2$.

The game associated to matrix (\ref{Matrix}) is an anticoordination game
because a cell obtains a larger payoff if it adopts the opposite strategy of
the other player. In other words, if the opponent cell adopts the
proliferative strategy, it is better for the cell to adopt the motile
strategy (and obtain $a-c$) than to proliferate (and obtain $b/2$).
Similarly, if the opponent cell adopts the motile strategy, it is better to
adopt the proliferative strategy (and obtain $b$) than to adopt the motile
strategy (and obtain $a/2-c/2$).

\begin{description}
\item[Example] Suppose the resource level to be shared by cells located in
the primary tumor is 3 ATP. In contrast, the resource level to be shared by
cells in a metastasis is 7 ATP. Let us say that the cost paid by cells to
move is 2 ATP. In this case we have $a=7$, $b=3$, $c=2$, and the game payoff
matrix can be set as: 
\begin{equation}
\begin{tabular}{c|cc}
& Motile & Proliferate \\ \hline
Motile & $2.5$ & $5$ \\ 
Proliferate & $3$ & $1.5$%
\end{tabular}
\label{Example 1 - Matrix}
\end{equation}%
If an opponent cell was to play the motile strategy, the best response of a
cell is to play the proliferative strategy obtaining a fitness of 3 instead
of 2.5. If an opponent cell was to play the proliferative strategy, the best
response of a cell would be to adopt the motile strategy, obtaining a
fitness of 5 instead of 1.5.
\end{description}

The different scenarios of competition between tumor cells are based on the
heterogeneity of the tumor. A punctuated-type tumor with low heterogeneity
is modeled by a game with only one type of player, while a branching-type
tumor with higher intratumor heterogeneity is modeled by a game with two
types of players. The concept of \textit{evolutionarily stable strategy}\
(ESS)\ introduced by Maynard Smith and Price \cite{Maynard-Price73} is
applied to solve both games. This notion captures the resilience of a given
strategy against any other strategy in the following sense. Consider a
population where most members play an evolutionarily stable strategy while a
small proportion of mutants choose a different strategy. In this situation
each mutant's expected payoff is smaller than the expected payoff of a
\textquotedblleft normal\textquotedblright\ individual, so that the mutants
are driven out from the population.

The remainder of the paper is organized as follows: Chapter 2 considers the
1-type game, modeling punctuated-type tumors in order to find out the
likelihood of a metastatic process. Chapter 3 extends the analysis to 2-type
games, modeling branching-type tumors. Chapter 4 compares both cases,
analyzing the difference in both cell motility and likelihood of metastasis.
Finally, Chapter 5 discusses the findings.

\section{Cellular motility in punctuated-type tumors}

As previously mentioned, \textit{BAP1}-mutated CCRCC are a paradigmatic
example of the punctuated-type tumor evolution in which the metastatic
competence and the clinical aggressiveness are high. In this tumor we model
cell motility by considering only pairwise meetings of \textit{BAP1}-type
cells, that is, \textit{BAP1}-\textit{BAP1} interactions.

In game theory, this corresponds to a go-or-grow population game with only
one type of player. Any pair of cells from the population of $n$ cells
plays the game represented by matrix (\ref{Matrix}). Let $\alpha $ denote
the probability of choosing the motile strategy\ so that a cell can choose
the pure strategy \textquotedblleft motile\textquotedblright\ ($\alpha =1$)
or to \textquotedblleft proliferate\textquotedblright\ ($\alpha =0$). A cell
can also adopt a mixed strategy ($0<\alpha <1$).

The evolutionarily stable strategy is the probability of adopting the motile
strategy which is stable in the sense defined above. The details of the
computation of the ESS are given in the Appendix. Here we give the intuition
of how we obtain it, where $\alpha $ is interpreted as the proportion of
motile cells in the population.

For a proportion $\alpha $ of motile cells, matrix (\ref{Matrix}) permits to
determine the average fitness of a proliferative cell (denoted by $%
v_{P}(\alpha )$) and a motile cell (denoted by $v_{M}(\alpha )$): 
\begin{eqnarray*}
v_{P}(\alpha ) &=&\alpha b+\frac{b}{2}\left( 1-\alpha \right) \text{ }=\frac{%
b}{2}\left( 1+\alpha \right) \text{ and} \\
v_{M}(\alpha ) &=&\alpha \frac{a-c}{2}+(1-\alpha )(a-c)=(a-c)\left( 1-\frac{%
\alpha }{2}\right) .
\end{eqnarray*}%
If a primary tumor is composed only of proliferative \textit{BAP1} cells
(i.e. $\alpha =0$), once a mutation introduces a single motile \textit{BAP1}
cell, the motile cell has a competitive advantage (that is, a larger
fitness: $a-c>b/2$), and thus can replicate at a faster rate than the
proliferative cells. This is true as long as $\alpha $ is sufficiently
small. Similarly if there were only motile cells (i.e. $\alpha =1$)
proliferative cells would have a competitive advantage (i.e. a larger
fitness: $b>a/2-c/2$). This is true for large values of $\alpha $. The
difference of fitness is given by: 
\begin{equation*}
v_{P}(\alpha )-v_{M}(\alpha )=\left( \frac{a+b-c}{2}\right) \alpha -\left( a-%
\frac{b}{2}-c\right) .
\end{equation*}%
When this difference is positive, motility to the metastatic tumor grows at
a slower rate than proliferation of cells in the primary tumor, driving the
proportion $\alpha $ of motile cells down. On the other hand, when this
difference is negative proliferation in the primary tumor grows at a slower
rate than motility to the metastatic tumor, driving the proportion $\alpha $
of motile cells up. When this difference is null both mutations lead to
the same cell fitness and cells replicate at the same growth rate in the
primary tumor. No mutation, motility or proliferation, has a competitive
advantage to the extent of driving the other mutation to extinction in the
primary tumor. The value of $\alpha $ that leads to a null difference
between the fitness of a proliferative cell and a motile cell is
evolutionarily stable. 
\begin{equation*}
v_{P}(\alpha )-v_{M}(\alpha )=0\text{ iff }\alpha =\left( a-\frac{b}{2}%
-c\right) /\left( \frac{a+b-c}{2}\right) .
\end{equation*}%
Formal proofs of the following results are given in the Appendix.

\textbf{Result 1: }The evolutionarily stable strategy in a punctuated-type
tumor associated to matrix (\ref{Matrix}) is given by\textbf{\ } 
\begin{equation}
\xi =\frac{2a-b-2c}{a+b-c}.  \label{ESS}
\end{equation}

Note that once metastasis develops, the metastatic and primary tumors
coexist (that is, $0<\xi <1$), independently of the cancer development
stage. The growth rate of each tumor depends on proliferation rate at the
primary tumor and both proliferation and motility to the metastatic tumor.
Proliferation rates in each tumor rely on the different levels of nutrients
available.

The evolutionarily stable strategy in a punctuated-type tumor can be
interpreted as the likelihood of a metastatic process, as $\xi $ gives the
probability of motility.

\textbf{Result 2:} All the rest being constant, the likelihood of a
metastatic process in punctuated-type tumors is larger (i) the larger
the amount of resource in the metastasis ($a$), (ii) the lower the amount of
resource in the primary tumor ($b$), and (iii) the lower the cost of moving (%
$c$).

The expected fitness of a cell, denoted by $v(\alpha )$, is given by%
\begin{eqnarray*}
v(\alpha ) &=&\alpha v_{M}(\alpha )+(1-\alpha )v_{P}(\alpha ) \\
&=&\frac{b}{2}+(a-c)\alpha -\frac{a+b-c}{2}\alpha ^{2}.
\end{eqnarray*}%
Substituting (\ref{ESS}) into the above equation we obtain the fitness
obtained by a cell at the ESS ($\alpha =\xi $): 
\begin{equation}
v(\xi )=\frac{b}{2}(1+\xi )  \label{v(pshi)}
\end{equation}

\textbf{Result 3:} At the ESS in punctuated-type tumors, the fitness
obtained by a cell lies between the fitness obtained if it was the only to
adopt the proliferative strategy and the fitness obtained if both cells
adopt the motile strategy. That is, 
\begin{equation*}
\frac{a-c}{2}<v(\xi )<b.
\end{equation*}

\begin{description}
\item[Example (continued)] The ESS in a punctuated-type tumor associated to
matrix (\ref{Example 1 - Matrix}) is given by: 
\begin{equation*}
\xi =\frac{14-3-4}{7+3-2}=\frac{7}{8}=0.875\text{ and }v(\xi )=\frac{3}{2}(1+%
\frac{7}{8})=\frac{45}{16}=2.8125.
\end{equation*}%
We have that $87.5$\% is the likelihood of a metastatic process and a cell
obtains a fitness of $2.8125$, with $2.5<2.8125<3$.
\end{description}

\section{Cellular motility in branching-type tumors}

Now, we consider a branching-type CCRCC, that is, a tumor with higher
intratumor heterogeneity in which multiple cell clones and subclones driven
by specific mutations develop during its temporal evolution. We will take
into consideration in the game only cells presenting two specific gene
mutations (apart from the \textit{VHL} gene mutation that is common to the
overwhelming majority of CCRCC). These mutations involve \textit{BAP1} and 
\textit{SETD2} genes and are associated with different metastatic
capability. So, \textit{BAP1} gene mutation is linked to high probability of
early multiple metastases (usually early and multiple) while \textit{SETD2}
gene mutation confers the tumor cell a low probability of metastases
(usually late and solitary). Given that both mutations provide CCRCC cells
with the metastatic competence, we model cell motility by considering the
possibility that not only cells with the same mutation can meet (\textit{BAP1%
}/\textit{BAP1} or \textit{SETD2}/\textit{SETD2}) but also that cells with
different mutations do it (\textit{BAP1}/\textit{SETD2} or \textit{SETD2}/%
\textit{BAP1}).

Interactions between cancer cells with different mutations can be regarded
as a game between players of different types. Types do not confer any
aprioristic advantage and the same payoff matrix (\ref{Matrix}) is played in
every encounter. Players do not know their own type, but detect the type of
their opponents. Examples of how cancer cells behave differently depending
on the recognition of their respective cellular contexts are available in
the literature \cite{Maruyama17}.

In game theory, this corresponds to a go-or-grow population game with two
types of players. For simplicity we name these types $i$-cells and $j$%
-cells. The respective proportions of $i$-cells and $j$-cells are
respectively denoted $x_{i}$ and $x_{j}$, with $x_{i}+x_{j}=1$. There are
four possible encounters between two cells: Both cells are $i$-cells; the
first cell is an $i$-cell and the second is a $j$-cell; the first is a $j$%
-cell and the second is an $i$-cell; or both are $j$-cells. Hence the
probabilities of these four possible encounters, which we denote
respectively by $p(i,i)$, $p(i,j)$, $p(j,i)$ and $p(j,j)$, are given by 
\begin{equation}
p(i,j)=p(j,i)=\frac{nx_{i}x_{j}}{n-1}\text{ for }i\neq j\text{ ; }p(i,i)=%
\frac{(nx_{i}-1)x_{i}}{n-1}\text{ and }p(j,j)=\frac{(nx_{j}-1)x_{j}}{n-1}.
\label{Probabilities}
\end{equation}

Any pair of cells from the population of $n$ cells plays the game
represented by matrix (\ref{Matrix}). In other words the types have no
intrinsic significance in the sense that any pair of cells plays the same
game (that is, faces the same payoff matrix described above).

As commented above we assume that cells do not know their type but recognize
the type of the opponent. That is, the behavior adopted by a cell does not
depend on its own type. By contrast, the behavior adopted may depend on the
type of the opponent. For instance, a cell may adopt a proliferative
behavior when meeting an $i$-cell and a motile behavior when meeting a $j$%
-cell.

A \textit{strategy} can thus be represented by $(\alpha _{i},\alpha _{j})$
where $\alpha _{i}$ gives the probability of adopting a motile behavior when
facing an $i$-cell and $\alpha _{j}$ gives the probability of adopting a
motile behavior when facing a $j$-cell. The expected payoff of a cell if all
cells play $(\alpha _{i},\alpha _{j})$ is denoted $V(\alpha _{i},\alpha
_{j}) $. We have 
\begin{equation*}
V(\alpha _{i},\alpha _{j})=p(i,i)u(\alpha _{i},\alpha _{i})+p(i,j)u(\alpha
_{j},\alpha _{i})+p(j,i)u(\alpha _{i},\alpha _{j})+p(j,j)u(\alpha
_{j},\alpha _{j})\text{,}
\end{equation*}%
where $u(\alpha ,\beta )$ is the expected fitness obtained by the cell which
moves with probability $\alpha $ while the opponent moves with probability $%
\beta $. See (\ref{u(alpha,beta)}) in the Appendix for details.

As previously for the 1-type game, we look for the ESS and the likelihood of
a metastatic process, i.e. the expected probability of being motile at the
ESS. If cells play strategy $(\alpha _{i},\alpha _{j})$, the expected
probability of being motile is the probability of being motile when meeting
an $i$-cell ($\alpha _{i}$) multiplied by the probability of meeting an $i$%
-cell ($x_{i}$) added to the probability of being motile when meeting a $j$%
-cell ($\alpha _{j}$) multiplied by the probability of meeting a $j$-cell ($%
x_{j}$). That is, it is given by $x_{i}\alpha _{i}+x_{j}\alpha _{j}$.

A first result is that the heterogeneity affects the likelihood of a
metastasis process. Indeed adopting the strategy that was stable in the
1-type game is not stable any more.

\textbf{Result 4: }$(\xi ,\xi )$ is not an evolutionarily stable strategy in
a branching-type tumor associated to matrix (\ref{Matrix}).

The strategies that are evolutionarily stable imply adopting different
behavior when facing different types of cells. If cells were motile with
probability $\xi $, cells using some different strategy could obtain a larger fitness and
finally invade the population. In the Appendix we exhibit a mutant strategy
that could invade a population playing $\xi $. However, there exist
strategies that cannot be invaded by a mutant strategy.

\textbf{Result 5: }Strategies $(\zeta _{i}^{\ast \ast },\zeta _{j}^{\ast
\ast })$ and $(\gamma _{i}^{\ast \ast },\gamma _{j}^{\ast \ast })$ are the
only evolutionarily stable strategies in a branching-type tumor associated
to matrix (\ref{Matrix}). These are respectively given by

\begin{eqnarray*}
(\zeta _{i}^{\ast \ast },\zeta _{j}^{\ast \ast }) &=&\left\{ 
\begin{array}{ll}
(\frac{n-1}{n-nx_{j}-1}\xi -\frac{nx_{j}}{n-nx_{j}-1},1) & \text{ if }x_{j}<%
\frac{n-1}{n}\xi \\ 
(0,1) & \text{if }\frac{n-1}{n}\xi <x_{j}<\frac{n-1}{n}\xi +\frac{1}{n} \\ 
(0,\frac{n-1}{nx_{j}-1}\xi ) & \text{if }x_{j}>\frac{n-1}{n}\xi +\frac{1}{n}%
\end{array}%
\right. \\
&&\text{and} \\
(\gamma _{i}^{\ast \ast },\gamma _{j}^{\ast \ast }) &=&\left\{ 
\begin{array}{cc}
(1,\frac{n-1}{n-nx_{i}-1}\xi -\frac{nx_{i}}{n-nx_{i}-1}) & \text{if }x_{i}<%
\frac{n-1}{n}\xi \\ 
(1,0) & \text{if }\frac{n-1}{n}\xi <x_{i}<\frac{n-1}{n}\xi +\frac{1}{n} \\ 
(\frac{n-1}{nx_{i}-1}\xi ,0) & \text{if }x_{i}>\frac{n-1}{n}\xi +\frac{1}{n}.%
\end{array}%
\right.
\end{eqnarray*}%
Note that the two ESS are the two faces of a same coin, the difference is
that for $(\zeta _{i}^{\ast \ast },\zeta _{j}^{\ast \ast })$ the cells that
induce the larger motility are the $j$-cells, while for $(\gamma _{i}^{\ast
\ast },\gamma _{j}^{\ast \ast })$ the cells that induce the larger motility
are the $i$-cells. In the following we focus on $(\zeta _{i}^{\ast \ast
},\zeta _{j}^{\ast \ast })$.

\begin{figure}[H]
	\centering
	\begin{tabular}{c}
		\epsfig{file=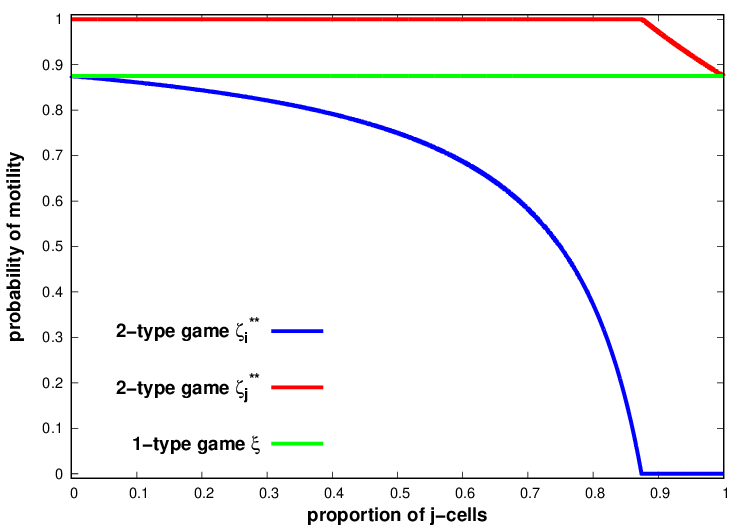,height=8cm,width=15cm,angle=0}
	\end{tabular}
	\vspace{-.15cm}  
	\caption{Probabilities of motility for each type of cell.}
	\label{fig:1}
\end{figure}

\begin{description}
\item[Example (continued)] Considering a population of $n=1001$ cells, the
ESS (where the $j$-cells induce more motility) in a branching-type tumor
associated to matrix (\ref{Example 1 - Matrix}) is given by: 
\begin{equation*}
(\zeta _{i}^{\ast \ast },\zeta _{j}^{\ast \ast })=\left\{ 
\begin{array}{ll}
(\frac{875-1001x_{j}}{1000-1001x_{j}},1) & \text{if }x_{j}<0.8741 \\ 
(0,1) & \text{if }0.8741<x_{j}<0.8751 \\ 
(0,\frac{875}{1001x_{j}-1}) & \text{if }x_{j}>0.8751.%
\end{array}%
\right.
\end{equation*}%
Figure \Ref{fig:1} represents $\zeta _{i}^{\ast \ast }$ and $\zeta _{j}^{\ast \ast }$
as functions of the proportion of $j$-cells and gives $\xi $ as a benchmark. 

It can be seen that the motility when facing an $i$-cell is smaller than the
motility in the punctuated-type tumor, which, in turn, is smaller than the
motility when facing a $j$-cell. For the extreme values of $x_{j}$ we
recover the results of the 1-type game: for $x_{j}=0$ we have $\zeta
_{i}^{\ast \ast }=\xi $ and for $x_{j}=1$ we have $\zeta _{j}^{\ast \ast
}=\xi $. When the proportion of $j$-cells increases the motility when facing
a $j$-cell decreases.
\end{description}

\section{Comparison of cellular motility}

Under which conditions the likelihood of a metastatic process is larger in a
punctuated-type tumor than in a branching-type tumor? The likelihood of a
metastatic process can be measured by the probability of adopting the motile
strategy. In a punctuated-type tumor, the evolutionarily stable strategy $%
\xi $ gives this probability.

In in a branching-type tumor the expected probability of being motile for a
cell is given by $x_{i}\alpha _{i}+x_{j}\alpha _{j}$. We can compute this
probability at the evolutionarily stable strategy and then compare the
probabilities. We obtain the following results (Proof in the Appendix).

\textbf{Result 6: }The probability of motility at the evolutionarily stable
strategy $(\zeta _{i}^{\ast \ast },\zeta _{j}^{\ast \ast })$ in
branching-type tumor associated to matrix (\ref{Matrix}) is given by 
\begin{equation*}
(1-x_{j})\zeta _{i}^{\ast \ast }+x_{j}\zeta _{j}^{\ast \ast }=\left\{ 
\begin{array}{ll}
\xi -\frac{x_{j}(1-\xi )}{n-nx_{j}-1} & \text{ if }x_{j}<\xi -\frac{1}{n}\xi
\\ 
x_{j} & \text{if }\xi -\frac{1}{n}\xi <x_{j}<\xi +(1-\xi )\frac{1}{n} \\ 
x_{j}\frac{n-1}{nx_{j}-1}\xi & \text{if }x_{j}>\xi +(1-\xi )\frac{1}{n}.%
\end{array}%
\right.
\end{equation*}

\textbf{Result 7: }We have: $(1-x_{j})\zeta _{i}^{\ast \ast }+x_{j}\zeta
_{j}^{\ast \ast }<\xi $ if and only if $x_{j}<\xi $ .

The likelihood of a metastatic process (i.e. the frequency of motility at
the evolutionarily stable strategy) is thus larger in a punctuated-type than
in a branching-type tumor if the proportion of the $j$-cells (those that induce
the larger motility) is not too large.

\begin{description}
\item[Example (continued)] Considering a population of $n=1001$ cells, the
expected motility (where the $j$-cells induce more motility) in a
branching-type tumor associated to matrix (\ref{Example 1 - Matrix}) is
given by 
\begin{equation*}
x_{i}\zeta _{i}^{\ast \ast }+x_{j}\zeta _{j}^{\ast \ast }=\left\{ 
\begin{array}{ll}
0.875-\frac{0.125x_{j}}{1000-1001x_{j}} & \text{if }x_{j}<0.8741 \\ 
x_{j} & \text{if }0.8741<x_{j}<0.8751 \\ 
\frac{875x_{j}}{1001x_{j}-1} & \text{if }x_{j}>0.8751.%
\end{array}%
\right.
\end{equation*}%
It is represented in Figure \ref{fig:2} and compared with the motility in a
punctuated-type tumor for different proportions of the $j$-cells. It can be
seen that for small proportions of $j$-cells, the motility is larger in a
punctuated-type tumor, while the reverse holds for large proportions of $j$%
-cells. Moreover a small increase in the proportion of $j$-cells has a
negative impact on the motility (with an exceptions for values\ of $x_{j}$
between $0.8741$ and $0.8751$).
\begin{figure}[H]
	\centering
	\begin{tabular}{c}
		\epsfig{file=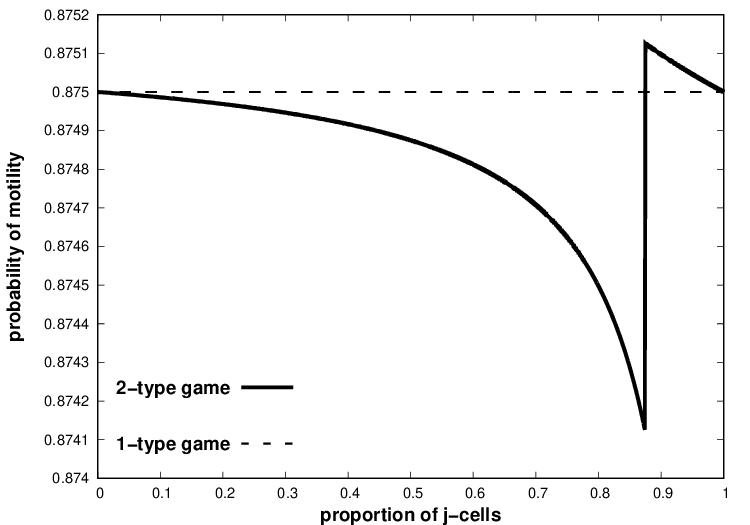,height=8cm,width=15cm,angle=0}
	\end{tabular}
	\vspace{-.15cm}  
	\caption{Probabilities of motility.}
	\label{fig:2}
\end{figure}
\end{description}

\section{Discussion}

We studied cell motility in CCRCC tumors using an evolutionary game theory
approach, in order to model intratumor cell interactions. Two key driver
mutations in such cancers were taken into account, \textit{BAP1} and \textit{%
SETD2}, which can be associated with different patterns of cancer
aggressiveness as well as with either punctuated or branching-type
evolutionary patterns.

The theoretical results found in the equilibrium in the mathematical model
support the empirical results observed in oncology, namely, the coexistence
of both primary and metastatic tumors and the conditions that favor a
metastatic process, particularly in the presence of resource scarcity in the
primary tumor and/or the presence of resource abundance in the metastatic
site. Favorable conditions for cell motility are also an incentive for
metastasis. Note that the level of cell heterogeneity may not be constant in
different areas of the same tumor leading to the so-called spatial
heterogeneity which follows ecological principles. In this sense, the tumor
interior, where the competition for the resources is fierce, the
heterogeneity is low because cells failing to get metastasizing competence
are destined to perish, whereas the tumor periphery, where resources and
oxygen are abundant, allow cells to develop different pathways of growth%
\textbf{\ }\cite{Zhao21}\textbf{.}

Moreover, in \cite{Laruelle23}, a game model gives mathematical support for
the higher aggressiveness empirically observed in punctuated cancers when
compared to branching ones in CCRCCs. Such result holds if the proportion of
the most aggressive cells remains below a given threshold. In this paper,
Result 7 provides additional support to such empirical observation given
that motility of cells is larger in punctuated cancers if the proportion of 
\textit{BAP1} cells remain below $\xi $.

\section*{Appendix}

\renewcommand{\thesection}{A.\arabic{section}} %
\setcounter{section}{0}

\renewcommand{\theequation}{A.\arabic{equation}} %
\setcounter{equation}{0}

\section{Fitness of a cell in the 1-type Go-or-Grow game}

In a bilateral encounter the payoff obtained by a cell depends on the
behavior of both cells. When the cell (resp. opponent) chooses to move with
probability $\alpha $ (resp. $\beta $), the expected fitness is 
\begin{eqnarray}
u(\alpha ,\beta ) &=&\alpha \beta \left( \frac{a-c}{2}\right) +\alpha
(1-\beta )(a-c)+(1-\alpha )\beta b+(1-\alpha )(1-\beta )\frac{b}{2}  \notag
\\
&=&\frac{b}{2}+\beta \frac{b}{2}+\left[ (a-\frac{b}{2}-c)-\beta \left( \frac{%
a+b-c}{2}\right) \right] \alpha .  \label{u(alpha,beta)}
\end{eqnarray}
The strategy that yields the highest payoff given that the opponent plays $%
\beta $ is $\alpha =1$ whenever $(a-\frac{b}{2}-c)-\beta \left( \frac{a+b-c}{%
2}\right) >0$, i.e. $\beta <\xi $. Similarly the optimal choice is $\alpha
=0 $ whenever $\beta >\xi $ and any $\alpha $ whenever $\beta =\xi $. That
is, the set of an individual's best responses to an opponent playing $\beta $%
, that we denote by $\mathcal{B}_{n}(\beta )$ is: 
\begin{equation}
\mathcal{B}_{n}(\beta )=\left\{ 
\begin{array}{ll}
\{1\} & \text{if }\beta <\xi \\ 
\{\alpha \left\vert \text{ }\alpha \in \lbrack 0,1]\right. \} & \text{if }%
\beta =\xi \\ 
\{0\} & \text{if }\beta >\xi .%
\end{array}%
\right.  \label{BestResponse}
\end{equation}%
An ESS (denoted $\alpha ^{\ast }$) must satisfy two conditions: \textit{(i)} 
$\alpha ^{\ast }$ has to be a best response to itself: $\alpha ^{\ast }\in 
\mathcal{B}(\alpha ^{\ast })$, and \textit{(ii)} if the opponent plays a
best response to $\alpha ^{\ast }$ then the payoff of playing $\alpha ^{\ast
}$ is strictly greater than the payoff of playing that best response.
Formally, for any $\beta \mathbf{\in }\mathcal{B}_{n}\mathbf{(}\alpha ^{\ast
}\mathbf{)}$ such that $\beta \neq \alpha $ we have $u(\alpha ^{\ast },\beta
)>u(\beta ,\beta ).$

\textbf{Result 1: }the ESS is given by\textbf{\ } 
\begin{equation*}
\xi =\frac{2a-b-2c}{a+b-c}.
\end{equation*}

\begin{proof}
It is easy to see from (\ref{BestResponse}) that strategy $\xi $ is the only
strategy that is a best response to itself, $\xi \in \mathcal{B}_{n}(\xi )$.
As any strategy is a best response to $\xi $ we check that the difference
between $u(\xi ,\beta )$ and $u(\beta ,\beta )$ is strictly positive:%
\begin{eqnarray*}
u(\xi ,\beta )-u(\beta ,\beta ) &=&\frac{b}{2}+\beta \frac{b}{2}-\xi \beta
\left( \frac{a+b-c}{2}\right) +\xi (a-\frac{b}{2}-c) \\
&&-\frac{b}{2}-\beta \frac{b}{2}+\beta \beta \left( \frac{a+b-c}{2}\right)
-\beta (a-\frac{b}{2}-c) \\
&=&\beta \left( \frac{a+b-c}{2}\right) (\beta -\xi )-(a-\frac{b}{2}-c)(\beta
-\xi ) \\
&=&(\beta -\xi )\frac{a+b-c}{2}(\beta -\frac{2a-b-2c}{a+b-c}) \\
&=&\frac{a+b-c}{2}(\beta -\xi )^{2}>0.
\end{eqnarray*}
\end{proof}

\textbf{Result 2:} (a) The larger the amount of resource in the metastasis,
all the rest constant, the larger the likelihood of a metastatic process is.
(b) The larger the amount of resource in the primary tumor, all the rest
constant, the lower the likelihood of a metastatic process is. (c) The
larger the cost of moving, all the rest constant, the lower the likelihood
of a metastatic process is.

\begin{proof}
Taking the first derivatives of the evolutionarily stable strategy with
respect to parameters a, b and c, we respectively obtain:%
\begin{eqnarray*}
\frac{\partial \xi }{\partial a} &=&\frac{3b}{(a-c+b)^{2}}>0 \\
\frac{\partial \xi }{\partial b} &=&\frac{-3(a-c)}{(a-c+b)^{2}}<0 \\
\frac{\partial \xi }{\partial c} &=&\frac{-3b}{(a-c+b)^{2}}<0.
\end{eqnarray*}
\end{proof}

\textbf{Result 3:} At the evolutionarily stable strategy, the fitness
obtained by a cell lies between the fitness obtained if it were the only to
adopt the proliferative strategy and the fitness obtained if both cells
adopt the motile strategy. That is, 
\begin{equation*}
\frac{a-c}{2}<v(\xi )<b.
\end{equation*}

\begin{proof}
Equation (\ref{v(pshi)}) gives 
\begin{equation*}
v(\xi )=\frac{b}{2}(1+\xi )=\frac{3}{2}\frac{b(a-c)}{a+b-c}.
\end{equation*}%
We then check that 
\begin{eqnarray*}
\frac{a-c}{2} &<&\frac{3}{2}\frac{b(a-c)}{a+b-c}\text{ given that }a+b-c<3b%
\text{ or }\frac{a-c}{2}<b. \\
\frac{3}{2}\frac{b(a-c)}{a+b-c} &<&b\text{ given that }3a-3c<2a+2b-2c\text{
or }\frac{a-c}{2}<b.
\end{eqnarray*}
\end{proof}

\section{Fitness of a cell in the 2-type Go-or-Grow game}

The expected payoff of an individual playing $(\alpha _{i},\alpha _{j})$
against an opponent playing $(\beta _{i},\beta _{j})$ is denoted by $%
U((\alpha _{i},\alpha _{j}),(\beta _{i},\beta _{j}))$. It is the sum of the
probability of each possible meeting multiplied by the expected utility in
each meeting. That is, 
\begin{equation*}
U((\alpha _{i},\alpha _{j}),(\beta _{i},\beta _{j}))=p(i,i)u(\alpha
_{i},\beta _{i})+p(i,j)u(\alpha _{j},\beta _{i})+p(j,i)u(\alpha _{i},\beta
_{j})+p(j,j)u(\alpha _{j},\beta _{j})
\end{equation*}%
Using (\ref{Matrix}) and (\ref{Probabilities}), we obtain 
\begin{eqnarray*}
&&U((\alpha _{i},\alpha _{j}),(\beta _{i},\beta _{j}))=\frac{b}{2}\left[
1+x_{i}\beta _{i}+x_{j}\beta _{j}\right] +\frac{a+b-c}{2}\left[ \xi -\frac{%
nx_{i}-1}{n-1}\beta _{i}-\frac{nx_{j}}{n-1}\beta _{j}\right] x_{i}\alpha _{i}
\\
&&+\frac{a+b-c}{2}\left[ \xi -\frac{nx_{i}}{n-1}\beta _{i}-\frac{nx_{j}-1}{%
n-1}\beta _{j}\right] x_{j}\alpha _{j}.
\end{eqnarray*}%
The expression above can be rewritten as 
\begin{equation}
U((\alpha _{i},\alpha _{j}),(\beta _{i},\beta _{j}))=f(\beta _{i},\beta
_{j})+f_{i}(\beta _{i},\beta _{j})\alpha _{i}+f_{j}(\beta _{i},\beta
_{j})\alpha _{j}  \label{U(alpha,beta)}
\end{equation}%
where 
\begin{eqnarray*}
f(\beta _{i},\beta _{j}) &=&\frac{b}{2}\left[ 1+x_{i}\beta _{i}+x_{j}\beta
_{j}\right] \\
f_{i}(\beta _{i},\beta _{j}) &=&\frac{a+b-c}{2}\left[ \xi -\frac{nx_{i}-1}{%
n-1}\beta _{i}-\frac{nx_{j}}{n-1}\beta _{j}\right] x_{i} \\
f_{j}(\beta _{i},\beta _{j}) &=&\frac{a+b-c}{2}\left[ \xi -\frac{nx_{i}}{n-1}%
\beta _{i}-\frac{nx_{j}-1}{n-1}\beta _{j}\right] x_{j}\text{.}
\end{eqnarray*}

It is easy to see that the optimal choice of an individual is $\alpha _{i}=1$
whenever $f_{i}(\beta _{i},\beta _{j})>0$, $\alpha _{i}=0$ whenever $%
f_{i}(\beta _{i},\beta _{j})<0$ and any $\alpha _{i}$ whenever $f_{i}(\beta
_{i},\beta _{j})=0$. Similarly the choice of $\alpha _{j}$ depends on the
sign of $f_{j}(\beta _{i},\beta _{j})$. Denoting by $\mathcal{B}%
_{n,x_{i},x_{j}}\mathbf{(}(\beta _{i},\beta _{j})\mathbf{)}$ the set of an
individual's best responses to an opponent playing $(\beta _{i},\beta _{j})$%
, we obtain: 
\begin{equation*}
\mathcal{B}_{n,x_{i},x_{j}}\mathbf{(}(\beta _{i},\beta _{j})\mathbf{)}%
=\left\{ 
\begin{array}{ll}
(\alpha _{i},\alpha _{j}) & \text{if }f_{i}(\beta _{i},\beta _{j})=0\text{
and }f_{j}(\beta _{i},\beta _{j})=0 \\ 
(\alpha _{i},1) & \text{if }f_{i}(\beta _{i},\beta _{j})=0\text{ and }%
f_{j}(\beta _{i},\beta _{j})>0 \\ 
(0,1) & \text{if }f_{i}(\beta _{i},\beta _{j})<0\text{ and }f_{j}(\beta
_{i},\beta _{j})>0 \\ 
(0,\alpha _{j}) & \text{if }f_{i}(\beta _{i},\beta _{j})<0\text{ and }%
f_{j}(\beta _{i},\beta _{j})=0 \\ 
(1,\alpha _{j}) & \text{if }f_{i}(\beta _{i},\beta _{j})>0\text{ and }%
f_{j}(\beta _{i},\beta _{j})=0 \\ 
(1,0) & \text{if }f_{i}(\beta _{i},\beta _{j})>0\text{ and }f_{j}(\beta
_{i},\beta _{j})<0 \\ 
(\alpha _{i},0) & \text{if }f_{i}(\beta _{i},\beta _{j})=0\text{ and }%
f_{j}(\beta _{i},\beta _{j})<0 \\ 
(1,1) & \text{if }f_{i}(\beta _{i},\beta _{j})>0\text{ and }f_{j}(\beta
_{i},\beta _{j})>0 \\ 
(0,0) & \text{if }f_{i}(\beta _{i},\beta _{j})<0\text{ and }f_{j}(\beta
_{i},\beta _{j})<0.%
\end{array}%
\right.
\end{equation*}%
An evolutionarily stable strategy must satisfy two conditions.

\begin{definition}
Strategy $(\alpha _{i}^{\ast \ast },\alpha _{j}^{\ast \ast })$ is \textit{%
evolutionarily stable} if and only if

\begin{enumerate}
\item The strategy has to be a best response to itself. That is, $(\alpha
_{i}^{\ast \ast },\alpha _{j}^{\ast \ast })\in \mathcal{B}_{n,x_{i},x_{j}}%
\mathbf{(}(\alpha _{i}^{\ast \ast },\alpha _{j}^{\ast \ast })\mathbf{)}$

\item If the opponent plays a best response to $(\alpha _{i}^{\ast \ast
},\alpha _{j}^{\ast \ast })$ then the payoff of playing $(\alpha _{i}^{\ast
\ast },\alpha _{j}^{\ast \ast })$ is strictly greater than the payoff of
playing that best response. Formally, for any $(\beta _{i},\beta _{j})%
\mathbf{\in }\mathcal{B}_{n,x_{i},x_{j}}\mathbf{(}(\alpha _{i}^{\ast \ast
},\alpha _{j}^{\ast \ast })\mathbf{)}$ such that $(\beta _{i},\beta
_{j})\neq (\alpha _{i}^{\ast \ast },\alpha _{j}^{\ast \ast })$ we have 
\begin{equation*}
U((\alpha _{i}^{\ast \ast },\alpha _{j}^{\ast \ast }),(\beta _{i},\beta
_{j}))>U((\beta _{i},\beta _{j}),(\beta _{i},\beta _{j})).
\end{equation*}
\end{enumerate}
\end{definition}

\textbf{Result 4: }$(\xi ,\xi )$ is not an evolutionarily stable strategy in
branching-type tumors.

\begin{proof}
This strategy is a best response to itself: $(\xi ,\xi )\in \mathcal{B}%
_{n,x_{i},x_{j}}(\xi ,\xi )$. Indeed we have that $f_{i}(\beta
_{i},\beta _{j})=0$ and $f_{j}(\beta _{i},\beta _{j})=0$ iff $\beta
_{i}=\beta _{j}=\xi $. Any strategy $(\beta _{i},\beta _{j})$ is a best
response to $(\xi ,\xi )$. We show that there exists a strategy $(\beta
_{i},\beta _{j})$ such that the difference between $U((\xi ,\xi ),(\beta
_{i},\beta _{j}))$ and $U((\beta _{i},\beta _{j}),(\beta _{i},\beta _{j}))\ $%
is negative. Using (\ref{U(alpha,beta)}) 
\begin{eqnarray*}
&&U((\xi ,\xi ),(\beta _{i},\beta _{j}))-U((\beta _{i},\beta _{j}),(\beta
_{i},\beta _{j})) \\
&=&f_{i}(\beta _{i},\beta _{j})(\xi -\beta _{i})+f_{j}(\beta _{i},\beta
_{j})(\xi -\beta _{j}) \\
&=&\frac{a+b-c}{2(n-1)}(\xi -\beta _{i})\left[ (n-1)\xi
x_{i}-x_{i}(nx_{i}-1)\beta _{i}-nx_{j}\beta _{j}x_{i}\right] \\
&&+\frac{a+b-c}{2(n-1)}(\xi -\beta _{j})\left[ (n-1)\xi x_{j}-nx_{i}\beta
_{i}x_{j}-(nx_{j}-1)\beta _{j}x_{j}\right] \\
&=&\frac{a+b-c}{2(n-1)}(\xi -\beta _{i})\left[ (n-1)\xi
x_{i}-x_{i}(nx_{i}-1)\beta _{i}-nx_{i}(1-x_{i})\beta _{j}\right] \\
&&+\frac{a+b-c}{2(n-1)}(\xi -\beta _{j})\left[ (n-1)\xi
(1-x_{i})-n(1-x_{i})x_{i}\beta _{i}-(n-nx_{i}-1)(1-x_{i})\beta _{j}\right] \\
&=&\frac{a+b-c}{2(n-1)}\left[ (\xi -\beta
_{i})x_{i}(nx_{i}-1)+nx_{i}(1-x_{i})\xi -nx_{i}(1-x_{i})\beta _{j}\right]
(\xi -\beta _{i}) \\
&&+\frac{a+b-c}{2(n-1)}\left[ (\xi -\beta
_{j})(1-x_{i})(n-nx_{i}-1)+nx_{i}(1-x_{i})\xi -nx_{i}(1-x_{i})\beta _{i}%
\right] (\xi -\beta _{j}) \\
&=&\frac{a+b-c}{2(n-1)}\left[ x_{i}(nx_{i}-1)(\xi -\beta
_{i})^{2}+nx_{i}(1-x_{i})\left( \xi -\beta _{j}\right) (\xi -\beta _{i})%
\right] \\
&&+\frac{a+b-c}{2(n-1)}\left[ (1-x_{i})(n-nx_{i}-1)(\xi -\beta
_{j})^{2}+nx_{i}(1-x_{i})\left( \xi -\beta _{i}\right) (\xi -\beta _{j})%
\right] \\
&=&\frac{a+b-c}{2(n-1)}\left[ x_{i}(nx_{i}-1)(\xi -\beta
_{i})^{2}+(1-x_{i})(n-nx_{i}-1)(\xi -\beta _{j})^{2}\right. \\
&&\left. +2nx_{i}(1-x_{i})\left( \xi -\beta _{i}\right) (\xi -\beta _{j}) 
\right]
\end{eqnarray*}

(i) If $n-nx_{i}-1=0$, we obtain%
\begin{eqnarray*}
&&U((\xi ,\xi ),(\beta _{i},\beta _{j}))-U((\beta _{i},\beta _{j}),(\beta
_{i},\beta _{j})) \\
&=&\frac{a+b-c}{2(n-1)}\left[ \frac{n-1}{n}(n-2)(\xi -\beta _{i})^{2}+2n%
\frac{n-1}{n}\left( \xi -\beta _{i}\right) (\xi -\beta _{j})\right] \\
&=&\frac{a+b-c}{2n}\left[ (n-2)(\xi -\beta _{i})+2(\xi -\beta _{j})\right]
\left( \xi -\beta _{i}\right) .
\end{eqnarray*}%
Choose $0<\beta _{i}<\xi $ and $\xi <\beta _{j}<1$ with $2\left( \beta
_{j}-\xi \right) >n-2$. Then 
\begin{equation*}
U((\xi ,\xi ),(\beta _{i},\beta _{j}))-U((\beta _{i},\beta _{j}),(\beta
_{i},\beta _{j}))<0.
\end{equation*}

(ii) If $n-nx_{i}-1\neq 0$ then let us rewrite%
\begin{eqnarray*}
&&U((\xi ,\xi ),(\beta _{i},\beta _{j}))-U((\beta _{i},\beta _{j}),(\beta
_{i},\beta _{j})) \\
&=&\frac{a+b-c}{2(n-1)}\left[ x_{i}(nx_{i}-1)(\xi -\beta
_{i})^{2}+(1-x_{i})(n-nx_{i}-1)(\xi -\beta _{j})^{2}\right. \\
&&\left. +2nx_{i}(1-x_{i})\left( \xi -\beta _{i}\right) (\xi -\beta _{j}) 
\right] \\
&=&\frac{a+b-c}{2(n-1)}(\beta _{i}-\xi )^{2}\left[
x_{i}(nx_{i}-1)+(1-x_{i})(n-nx_{i}-1)(\frac{\xi -\beta _{j}}{\beta _{i}-\xi }%
)^{2}\right. \\
&&\left. -2nx_{i}(1-x_{i})\frac{\xi -\beta _{j}}{\beta _{i}-\xi }\right]
\end{eqnarray*}

Denote $Z=\frac{\xi -\beta _{j}}{\beta _{i}-\xi }$ we obtain%
\begin{eqnarray*}
&&U((\xi ,\xi ),(\beta _{i},\beta _{j}))-U((\beta _{i},\beta _{j}),(\beta
_{i},\beta _{j})) \\
&=&\frac{a+b-c}{2(n-1)}(\beta _{i}-\xi )^{2}\left[
(1-x_{i})(n-nx_{i}-1)(Z)^{2}-2nx_{i}(1-x_{i})Z+x_{i}(nx_{i}-1)\right] 
\end{eqnarray*}%
The discriminant is 
\begin{eqnarray*}
\Delta  &=&4n^{2}x_{i}^{2}(1-x_{i})^{2}-4x_{i}(1-x_{i})(n-nx_{i}-1)(nx_{i}-1)
\\
&=&4x_{i}(1-x_{i})\left[
n^{2}x_{i}-n^{2}x_{i}^{2}-n^{2}x_{i}+n^{2}x_{i}^{2}+nx_{i}+n-nx_{i}-1\right] 
\\
&=&4x_{i}(1-x_{i})(n-1)>0
\end{eqnarray*}%
It is strictly positive.\footnote{%
Note that if $x_{i}=0$ or $x_{i}=1$ twe do not have 2-types of players any
more. Here we exclude these possibilities.} This means that the difference $%
U((\xi ,\xi ),(\beta _{i},\beta _{j}))-U((\beta _{i},\beta _{j}),(\beta
_{i},\beta _{j}))$ will take negative values for $\frac{2nx_{i}(1-x_{i})-%
\sqrt{\Delta }}{2(1-x_{i})(n-nx_{i}-1)}<Z<\frac{2nx_{i}(1-x_{i})+\sqrt{%
\Delta }}{2(1-x_{i})(n-nx_{i}-1)}$. In particular choosing $Z=\frac{%
2nx_{i}(1-x_{i})}{2(1-x_{i})(n-nx_{i}-1)}=\frac{nx_{i}}{n-nx_{i}-1}$ we
obtain%
\begin{eqnarray*}
&&U((\xi ,\xi ),(\beta _{i},\beta _{j}))-U((\beta _{i},\beta _{j}),(\beta
_{i},\beta _{j})) \\
&=&\frac{a+b-c}{2(n-1)}(\beta _{i}-\xi )^{2}\left[ (1-x_{i})(n-nx_{i}-1)(%
\frac{nx_{i}}{n-nx_{i}-1})^{2}\right.  \\
&&\left. -2nx_{i}(1-x_{i})\frac{nx_{i}}{n-nx_{i}-1}+x_{i}(nx_{i}-1)\right] 
\\
&=&\frac{a+b-c}{2(n-1)}(\beta _{i}-\xi )^{2}\frac{1}{n-nx_{i}-1}\left[
(1-x_{i})(nx_{i})^{2}-2(nx_{i})^{2}(1-x_{i})\right.  \\
&&\left. +x_{i}(nx_{i}-1)(n-nx_{i}-1)\right]  \\
&=&\frac{a+b-c}{2(n-1)}(\beta _{i}-\xi )^{2}\frac{1}{n-nx_{i}-1}\left[
n^{2}x_{i}^{2}-n^{2}x_{i}^{3}-2n^{2}x_{i}^{2}+2n^{2}x_{i}^{3}+n^{2}x_{i}^{2}-n^{2}x_{i}^{3}\right. 
\\
&&\left. -nx_{i}-n+nx_{i}+1\right]  \\
&=&\frac{a+b-c}{2(n-1)}(\beta _{i}-\xi )^{2}\frac{1}{n-nx_{i}-1}\left[ -n+1%
\right] <0.
\end{eqnarray*}
\end{proof}

\textbf{Result 5: }Strategies $(\zeta _{i}^{\ast \ast },\zeta _{j}^{\ast
\ast })$ and $(\gamma _{i}^{\ast \ast },\gamma _{j}^{\ast \ast })$ are the
only evolutionarily stable strategies.

\begin{proof}[Proof]
We can check that $(\zeta _{i}^{\ast \ast },\zeta _{j}^{\ast \ast })$
satisfies the two conditions for being an ESS:

\begin{enumerate}
\item $(\zeta _{i}^{\ast \ast },\zeta _{j}^{\ast \ast })\in \mathcal{B}%
_{n,x_{i},x_{j}}(\zeta _{i}^{\ast \ast },\zeta _{j}^{\ast \ast })$. We have
that $f_{i}((\beta _{i},1))=0$ iff $\beta _{i}=\frac{n-1}{n-nx_{j}-1}\xi -%
\frac{nx_{j}}{n-nx_{j}-1}$. The condition $x_{j}<\frac{n-1}{n}\xi $
guarantees that $\beta _{i}>0$ and $f_{j}((\beta _{i},1))>0$. Second $%
f_{i}((0,1))<0$ and $f_{j}((0,1))>0$ iff $\frac{n-1}{n}\xi <x_{j}<\frac{n-1}{%
n}\xi +\frac{1}{n}$. Third, we have $f_{j}(0,\beta _{j})=0$ iff $\beta _{j}=%
\frac{n-1}{nx_{j}-1}\xi $. Condition $x_{j}>\frac{n-1}{n}\xi +\frac{1}{n}$
guarantees that $\beta _{j}<1$ and $f_{i}(0,\beta _{j})<0$.

\item The set of an individual's best responses to an opponent playing $%
(\zeta _{i}^{\ast \ast },\zeta _{j}^{\ast \ast })$ is 
\begin{equation*}
\mathcal{B}_{n,x}(\zeta _{i}^{\ast \ast },\zeta _{j}^{\ast \ast })=\left\{ 
\begin{array}{ll}
\{(\beta _{i},1)\left\vert \text{ }\beta _{i}\in \left[ 0,1\right] \right. \}
& \text{if }x_{j}<\frac{n-1}{n}\xi \\ 
\{(0,1)\} & \text{if }\frac{n-1}{n}\xi <x_{j}<\frac{n-1}{n}\xi +\frac{1}{n}
\\ 
\{(0,\beta _{j})\left\vert \text{ }\beta _{j}\in \left[ 0,1\right] \right. \}
& \text{if }x_{j}>\frac{n-1}{n}\xi +\frac{1}{n}.%
\end{array}%
\right.
\end{equation*}%
For $\frac{n-1}{n}\xi <x_{j}<\frac{n-1}{n}\xi +\frac{1}{n}$ we have that $%
(0,1)$ is the \textit{only} best response to itself. For $x_{j}<\frac{n-1}{n}%
\xi $ we have: 
\begin{eqnarray*}
&&U((\frac{n-1}{n-nx_{j}-1}\xi -\frac{nx_{j}}{n-nx_{j}-1},1),(\beta
_{i},1))-U((\beta _{i},1),(\beta _{i},1)) \\
&=&f_{i}(\beta _{i},\beta _{j})(\frac{n-1}{n-nx_{j}-1}\xi -\frac{nx_{j}}{%
n-nx_{j}-1}-\beta _{i}) \\
&=&\frac{a+b-c}{2(n-1)}(1-x_{j})\left[ (n-1)\xi -(n-nx_{j}-1)\beta
_{i}-nx_{j}\right] \left[ \frac{n-1}{n-nx_{j}-1}\xi \right. \\
&&\left. -\frac{nx_{j}}{n-nx_{j}-1}-\beta _{i}\right] \\
&=&\frac{a+b-c}{2(n-1)(n-nx_{j}-1)}(1-x_{j})\left[ (n-1)\xi
-(n-nx_{i}-1)\beta _{i}-nx_{j})\right] \left[ (n-1)\xi \right. \\
&&\left. -nx_{j}-(n-nx_{i}-1)\beta _{i}\right] \\
&=&\frac{a+b-c}{2(n-1)(n-nx_{j}-1)}(1-x_{j})\left[ (n-1)\xi
-(n-nx_{i}-1)\beta _{i}-nx_{j})\right] ^{2}>0.
\end{eqnarray*}%
For $x_{j}>\frac{n-1}{n}\xi +\frac{1}{n}$ we have: 
\begin{eqnarray*}
&&U((0,\frac{n-1}{nx_{j}-1}\xi ),(0,\beta _{j}))-U((0,\beta _{j}),(0,\beta
_{j})) \\
&=&f_{j}(\beta _{i},\beta _{j})(\frac{n-1}{nx_{j}-1}\xi -\beta _{j}) \\
&=&\frac{a+b-c}{2(n-1)}\left[ (n-1)\xi x_{j}-(nx_{j}-1)\beta _{j}x_{j}\right]
(\frac{n-1}{nx_{j}-1}\xi -\beta _{j}) \\
&=&\frac{a+b-c}{2(n-1)(nx_{j}-1)}x_{j}\left[ (n-1)\xi -(nx_{j}-1)\beta _{j}%
\right] \left[ (n-1)\xi -(nx_{j}-1)\beta _{j}\right] \\
&=&\frac{a+b-c}{2(n-1)(nx_{j}-1)}x_{j}\left[ (n-1)\xi -(nx_{j}-1)\beta _{j}%
\right] ^{2}>0.
\end{eqnarray*}
\end{enumerate}

The proof that strategy $(\gamma _{i}^{\ast \ast },\gamma _{j}^{\ast \ast })$
is evolutionarily stable is omitted because it is similar to the previous
proof. These are the only ESS. Result 4 has shown that $(\xi ,\xi )$ is not
an ESS. No other strategy is a best response to itself given that $%
f_{i}((1,1))<0$, and $f_{i}((0,0))>0$.
\end{proof}

\section{Comparison of cellular motility}

\textbf{Result 6: }The probability of motility at the evolutionarily stable
strategy $(\zeta _{i}^{\ast \ast },\zeta _{j}^{\ast \ast })$ in
branching-type tumor is given by 
\begin{equation*}
(1-x_{j})\zeta _{i}^{\ast \ast }+x_{j}\zeta _{j}^{\ast \ast }=\left\{ 
\begin{array}{ll}
\xi -\frac{x_{j}(1-\xi )}{n-nx_{j}-1} & \text{ if }x_{j}<\xi -\frac{1}{n}\xi
\\ 
x_{j} & \text{if }\xi -\frac{1}{n}\xi <x_{j}<\xi +(1-\xi )\frac{1}{n} \\ 
x_{j}\frac{n-1}{nx_{j}-1}\xi & \text{if }x_{j}>\xi +(1-\xi )\frac{1}{n}.%
\end{array}%
\right.
\end{equation*}

\begin{proof}[Proof]
(i) if $x_{j}<\xi -\frac{1}{n}\xi $ we have%
\begin{equation*}
\begin{tabular}{ll}
$x_{i}\zeta _{i}^{\ast \ast }+x_{j}\zeta _{j}^{\ast \ast }$ & $=x_{i}\left( 
\frac{n-1}{n-nx_{j}-1}\xi -\frac{nx_{j}}{n-nx_{j}-1}\right) +x_{j}1$ \\ 
& $=\frac{(1-x_{j})(n-1)\xi -nx_{j}(1-x_{j})+\left( n-nx_{j}-1\right) x_{j}}{%
n-nx_{j}-1}$ \\ 
& $=\frac{(n-nx_{j}-1)\xi +x_{j}\xi -x_{j}}{n-nx_{j}-1}$ \\ 
& $=\xi -\frac{x_{j}(1-\xi )}{n-nx_{j}-1};$%
\end{tabular}%
\end{equation*}%
(ii) if $\xi -\frac{1}{n}\xi <x_{j}<\xi +(1-\xi )\frac{1}{n}$ we have $%
x_{i}\zeta _{i}^{\ast \ast }+x_{j}\zeta _{j}^{\ast \ast }=x_{j}$; and (iii)
if $x_{j}>\frac{n-1}{n}\xi +\frac{1}{n}$ we have $x_{i}\zeta _{i}^{\ast \ast
}+x_{j}\zeta _{j}^{\ast \ast }=x_{j}\frac{n-1}{nx_{j}-1}\xi .$
\end{proof}

\textbf{Result 7: }We have: $(1-x_{j})\zeta _{i}^{\ast \ast }+x_{j}\zeta
_{j}^{\ast \ast }<\xi $ if and only if $x_{j}<\xi $ .

\begin{proof}[Proof]
(i) if $x_{j}<\xi -\frac{1}{n}\xi $ we have $x_{i}\zeta _{i}^{\ast \ast
}+x_{j}\zeta _{j}^{\ast \ast }=\xi -\frac{x_{j}(1-\xi )}{n-nx_{j}-1}<\xi $;
(ii) if $\xi -\frac{1}{n}\xi <x_{j}<\xi +(1-\xi )\frac{1}{n}$ we have $%
x_{i}\zeta _{i}^{\ast \ast }+x_{j}\zeta _{j}^{\ast \ast }=x_{j}$; and (iii)
if $x_{j}>\xi +(1-\xi )\frac{1}{n}$ we have $x_{i}\zeta _{i}^{\ast \ast
}+x_{j}\zeta _{j}^{\ast \ast }=\frac{nx_{j}-x_{j}}{nx_{j}-1}\xi >\xi $,
given that $nx_{j}-x_{j}>nx_{j}-1.$
\end{proof}

\section*{Acknowledgments}

\noindent Rocha acknowledges financial
support from the National Council for Scientific and Technological 
Development – CNPq (CNPq funding 307437/2019-1); 
Laruelle acknowledges financial support from grant
PID2019-106146GB-I00 funded by MCIN/AEI/
10.13039/ 501100011033 and by
\textquotedblleft ERDF, A way of making Europe\textquotedblright\ and from
the Basque Government (Research Group IT1697-22).


\end{document}